\documentclass[a4paper]{PoS}

\usepackage[utf8]{inputenc}
\usepackage[T1]{fontenc}
\usepackage[english]{babel}
\usepackage{amsmath}
\usepackage{amsfonts}
\usepackage{cite}

\def\beq{\begin{equation}}
\def\eeq{\end{equation}}
\def\bsp#1\esp{\begin{split}#1\end{split}}
\def\bal#1\eal{\begin{align}#1\end{align}}

\newcommand\dd {\mathrm{d}}
\newcommand\ordo[1] {\mathcal{O}(#1)}

\title{A new formulation of the loop-tree duality at higher loops}

\ShortTitle{A new formulation of the loop-tree duality at higher loops}

\author{Robert Runkel\\
        PRISMA Cluster of Excellence, Institut f{\"u}r Physik, 
        Johannes Gutenberg-Universit{\"a}t Mainz,\\
        D - 55099 Mainz, Germany\\
        E-mail: \email{rrunkel@uni-mainz.de}}

\author{\speaker{Zolt\'an Sz\H{o}r}\\
        PRISMA Cluster of Excellence, Institut f{\"u}r Physik, 
        Johannes Gutenberg-Universit{\"a}t Mainz,\\
        D - 55099 Mainz, Germany\\
        E-mail: \email{zoltanszoer@uni-mainz.de}}

\author{Juan Pablo Vesga\\
        PRISMA Cluster of Excellence, Institut f{\"u}r Physik, 
        Johannes Gutenberg-Universit{\"a}t Mainz,\\
        D - 55099 Mainz, Germany\\
        E-mail: \email{juvesgas@uni-mainz.de}}

\author{Stefan Weinzierl\\
        PRISMA Cluster of Excellence, Institut f{\"u}r Physik, 
        Johannes Gutenberg-Universit{\"a}t Mainz,\\
        D - 55099 Mainz, Germany\\
        E-mail: \email{weinz001@uni-mainz.de}}

\abstract{
We present a new formulation of the loop-tree duality theorem for higher loop diagrams 
valid both for massless and massive cases. 
$l$-loop integrals are expressed as weighted sum of trees obtained from cutting $l$ internal 
propagators of the loop graph. 
In addition, the uncut propagators gain a modified $i \delta$-prescription, named dual-propagators.
In this new framework one can go beyond graphs 
and calculate the integrand of loop amplitudes as a weighted sum of tree graphs, which form a 
tree-like object. 
These objects can be computed efficiently via recurrence relations.}

\FullConference{14th International Symposium on Radiative Corrections (RADCOR2019)\\ 
9-13 September 2019\\
		Palais des Papes, Avignon, France}

\begin{document}

%------------------------------------
%           Introduction
%------------------------------------
%
\section{Introduction}
\label{sec:intro}
The precision physics of the LHC requires accurate predictions, 
especially with the upcoming high luminosity era. 
One of the bottlenecks of higher order predictions in perturbation theory is the 
computation of loop amplitudes beyond one-loop order.
If one goes beyond the simplest processes involving $2 \rightarrow 2$ scatterings, general methods  
which allow for high level of automation are preferred.
The loop-tree duality approach \cite{Catani:2008xa,Bierenbaum:2010cy,Bierenbaum:2012th,Buchta:2014dfa,
Hernandez-Pinto:2015ysa,Buchta:2015wna,Sborlini:2016gbr,Driencourt-Mangin:2017gop,Driencourt-Mangin:2019aix,
Aguilera-Verdugo:2019kbz,Runkel:2019yrs,Baumeister:2019rmh,Runkel:2019zbm} 
combined with numerical loop integration techniques \cite{Soper:1998ye,Soper:1999xk,Nagy:2003qn,Gong:2008ww,
Assadsolimani:2009cz,Assadsolimani:2010ka,Becker:2010ng,Becker:2011vg,Becker:2012aqa,Becker:2012nk,Becker:2012bi,
Goetz:2014lla,Seth:2016hmv,Anastasiou:2018rib}
can serve as basis for such an automated framework.

Relating loop integrals to trees dates back to Feynman \cite{Feynman:1963ax}, where the loop integral 
is expressed as a sum of forests.
Inspired by this idea loop-tree duality was formulated at one-loop order nearly a decade ago \cite{Catani:2008xa}.
In this approach loop graphs can be expressed as sum of tree graphs. 
Propagators inherit a modified $i \delta$ prescription, hence they are called dual-propagators.
Dual-propagators exhibit the cancellations of unphysical singularities present on so-called H-surfaces 
\cite{Buchta:2014dfa}, simplifying the singularity structure of the integrand.
Furthermore at next-to-leading order the infrared structure of virtual correction can be matched 
with the infrared structure of the real term leading to infrared cancellation on the integrand level  
\cite{Hernandez-Pinto:2015ysa,Sborlini:2016gbr,Driencourt-Mangin:2017gop}.

From two-loops onwards there are two different ways of progress: one is based on distributional identities 
between Feynman- and dual-propagators \cite{Bierenbaum:2010cy}, while the other is based on taking the 
$l$-fold residue \cite{Runkel:2019yrs,Capatti:2019ypt}.
The two different ways also lead to different modified $i \delta$ prescriptions.
Dual cancellations carry on to higher loop orders as well 
\cite{Driencourt-Mangin:2019aix,Aguilera-Verdugo:2019kbz,Capatti:2019ypt}, 
but the local cancellation of infrared singularities amongst different corrections is still part of ongoing research.

In this talk we present a formulation of the loop-tree duality at higher loop orders, which is based 
on the calculation of residues. 
First we discuss the method and its features for multiloop graphs in Sect. \ref{sec:ltd-graphs}, 
then we move to loop amplitudes in Sect. \ref{sec:ltd-amps}.
Our final aim is to provide a general automated framework to compute loop amplitudes beyond one-loop 
and achive local infrared cancellation, hence we construct our method keeping this as leading principle.

%------------------------------------
%    Loop-tree duality for graphs  
%------------------------------------
%
\section{Loop-tree duality for graphs}
\label{sec:ltd-graphs}
We start with the discussion of our loop-tree duality approach for graphs.
We consider a general $l$-loop graph $\Gamma$ with $n$ external legs.
The corresponding integral is 
\beq
\label{eq:intdef}
I = \int \prod_{i=1}^l \frac{\dd^D k_i}{(2 \pi)^D} \frac{P_{\Gamma}}{\prod\limits_{j} 
\big(k_j^2 - m_j^2 + i \delta \big)}
= \int \prod_{i=1}^l \frac{\dd^D k_i}{(2 \pi)^D} f(\Gamma) \,,
\eeq
where $k_j$ is the momenta flowing through propagator $j$ and is in general a linear combination 
of external and loop momenta. 
For the integrand we introduced the short-hand notation
\beq
\label{eq:fdef}
f(\Gamma) = \frac{P_{\Gamma}}{\prod\limits_{j} \big(k_j^2 - m_j^2 + i \delta \big)} \,,
\eeq
where the numerator $P_{\Gamma}$ is a polynomial in both loop and external momenta.
Furthermore we assume that the form of $P_{\Gamma}$ is such that 
all energy integrations over half circles at infinity vanish.
At $l$-loops we can perform the energy integration of each loop with the help of the residue theorem,  
and the loop integral is given by the weighted sum of $l$-fold residues 
\beq
\label{eq:ltd-graphs}
\int \prod\limits_{i=1}^l \frac{\dd^D k_i}{(2 \pi)^D} f(\Gamma)
= (-i)^l \sum\limits_{\sigma \in C_{\Gamma}} \int\limits_{+/-} \prod\limits_{i=1}^l
\frac{\dd^{D-1} k_{\sigma_i}}{(2 \pi)^{D-1}}
S_{\sigma \alpha} (-1)^{n_{\sigma}^{(\alpha)}} \text{res} \Big(f,E_{\sigma}^{(\alpha)}\Big) \,.
\eeq
The notation is the following: $\sigma$ denotes the set of indices of cutting $l$ 
edges of the graph $\Gamma$ such that the obtained graph is a connected tree graph.
The set of all such cuts is denoted by $C_{\Gamma}$. \\
The $l$-tuples 
\beq
E_{\sigma}^{(\alpha)} = \bigg(\pm \sqrt{\vec{k}^2_{\sigma_1}+m^2_{\sigma_1} - i \delta}, \dots,  
\pm \sqrt{\vec{k}^2_{\sigma_l}+m^2_{\sigma_l} - i \delta}
\bigg)
\eeq
with $\alpha \in \{-1,1\}^l$ are the solutions to the equations
\beq
\label{eq:prop-onshell}
k^2_{\sigma_1} - m^2_{\sigma_1} + i \delta = \dots =
k^2_{\sigma_l} - m^2_{\sigma_l} + i \delta = 0 \,,
\eeq
and $n_{\sigma}^{(\alpha)}$ the number of positive components in the solution. 
The subscript of the integration in Eq. (\ref{eq:ltd-graphs}) indicates an implicit summation 
over $\alpha$, namely
\beq
\int\limits_{+/-} \frac{\dd^{D-1} k_{\sigma_i}}{(2 \pi)^{D-1}} f(k_{\sigma_i}) 
= \int \frac{\dd^{D-1} k_{\sigma_i}}{(2 \pi)^{D-1}} 
\bigg[ f\bigg(+ \sqrt{\vec{k}^2_{\sigma_i}+m^2_{\sigma_i}}, \vec{k}_{\sigma_i} \bigg) 
+ f\bigg(- \sqrt{\vec{k}^2_{\sigma_i}+m^2_{\sigma_i}}, \vec{k}_{\sigma_i} \bigg) \bigg] \,.
\eeq
Finally $S_{\sigma \alpha}$ are combinatorial factors, which we will discuss later in more detail.

When all propagators appear to power one, then the residues in Eq. (\ref{eq:ltd-graphs}) are given by 
\beq
\label{eq:residues}
(-1)^{n_{\sigma}^{(\alpha)}} \text{res} \Big(f,E_{\sigma}^{(\alpha)}\Big) = 
\prod\limits_{i=1}^l \frac{1}{2\sqrt{\vec{k}^2_{\sigma_i}+m^2_{\sigma_i}}} 
\frac{P_{\Gamma}}{\prod\limits_{j \notin \sigma} \big(k_j^2 - m_j^2 + i\delta s_j(\sigma) \big)}
\eeq
that is a product of the uncut propagators.
We call these propagators dual-propagators, since they inherit a modified $i \delta$ prescription. 
This information is encoded in $s_j(\sigma)$, which can be computed systematically for each 
dual-propagator \cite{Runkel:2019yrs}.
The product of dual propagators can be interpreted as a tree graph; we deleted $l$ propagators from our 
loop graph in such a way, that the resulting graph is a connected tree graph.
The structure of this connected tree graph is encoded in Eq. (\ref{eq:residues}).
Hence we can truly name Eq. (\ref{eq:ltd-graphs}) loop-tree duality, as it relates a $D$-dimensional loop integral 
to the $D-1$ dimensional phase space integral of weighted sum of tree graphs.

Now we discuss the combinatorial factors $S_{\sigma \alpha}$ appearing in Eq. (\ref{eq:ltd-graphs}) 
in more detail, as they are an important feature of our method.
In our approach we sum over all possible cuts $\sigma$, all possible energy configurations $\alpha$, 
and all orders of picking up the residues $\pi$.
However the representation of the integral is not unique in terms of cuts \cite{Capatti:2019ypt}.
Formally we can write the decomposition in terms of cuts as 
\beq
I = \int \sum_{\sigma \in C_{\Gamma}} \sum_{\pi \in S_l} \sum_{\alpha \in \{-1,1\}^l} 
C_{\sigma \pi \alpha}^{\tilde{\sigma} \tilde{\pi} \tilde{\alpha}} \text{Cut}(\sigma,\alpha) \,,
\eeq
where $C_{\sigma \pi \alpha}^{\tilde{\sigma} \tilde{\pi} \tilde{\alpha}}$ is the coefficient of 
a certain cut.
This coefficient depends on subjective choices.
We have the freedom to assign the loop momenta to $l$ propagators, which is denoted by $\tilde{\sigma}$. 
The order of integration is also subjective, we label it by $\tilde{\pi}$.
At last we can close the integration contour on the upper or the lower complex plane 
for each loop integration separately, this choice is given by $\tilde{\alpha}$.
Each choice is equivalent to any other and provides the same integral, however the representation is different.
In order for the result be independent of these arbirtrary choices we average over $\tilde{\sigma}$, $\tilde{\pi}$ 
and $\tilde{\alpha}$ and sum over $\pi$.
Then the combinatorial factors are given by
\beq
\label{eq:combdef}
S_{\sigma \alpha} = \frac{(-1)^{l+n_{\sigma}^{(\alpha)}}}{2^l l! |C_{\Gamma^{\text{chain}}}|}
\sum_{\pi \in S_l} \sum_{\tilde{\sigma} \in C_{\Gamma}} \sum_{\tilde{\pi} \in S_l} \sum_{\tilde{\alpha} \in \{-1,1\}^l}
\frac{C_{\sigma \pi \alpha}^{\tilde{\sigma} \tilde{\pi} \tilde{\alpha}}}{N^{\text{chain}}(\sigma)} \,,
\eeq
where we use the concept of chain graphs \cite{Kinoshita:1962ur}. 
Two propagators are grouped to the same chain, if their momenta differ only by a linear combination 
of the external momenta.
$N^{\text{chain}}$ is given by the product of the number of propagators present in each chain, 
while $|C_{\Gamma^{\text{chain}}}|$ denotes the number of ways a chain graph can be cut into a 
connected tree graph.
Eq. (\ref{eq:combdef}) provides a definition of the combinatorial factors, which is independent on the number 
of external legs, it only depends on the underlying chain graph.
Hence $S_{\sigma \alpha}$ must be computed only once for a given chain graph.
For example, at two-loops the only chain graph is the sunrise graph shown as the first graph of Table \ref{tab:graphs}.
The corresponding combinatorial factors are presented in Table \ref{tab:combfact}. 
\begin{table}[ht]
\centering
\begin{tabular}{c|c|c}
Cut & $(1^+,2^+)$ & $(1^+,2^-)$ \\
\hline
$S_{\sigma \alpha}$ & $1/3$ & $1/6$ 
\end{tabular}
\caption{Combinatorial factors for the two-loop sunrise chain graph shown in Table \ref{tab:graphs}.}
\label{tab:combfact}
\end{table}

Next we consider four examples pictured in Table \ref{tab:graphs}.
\begin{table}[ht]
\centering
\begin{tabular}{c c c c}
Sunrise & Mercedes & Fly & N.p. double box \\
\includegraphics[scale=0.65]{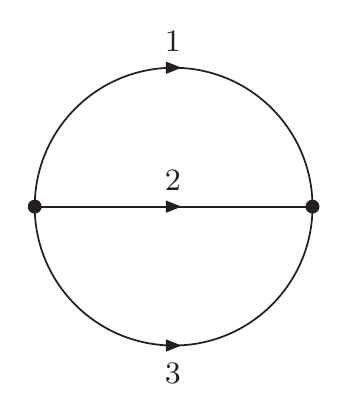} &
\includegraphics[scale=0.65]{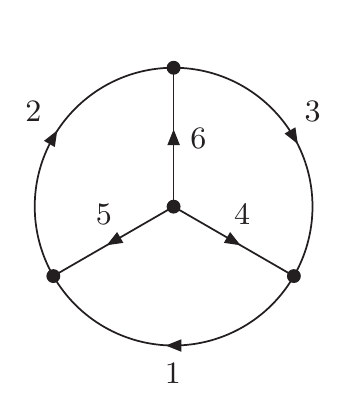} &
\includegraphics[scale=0.65]{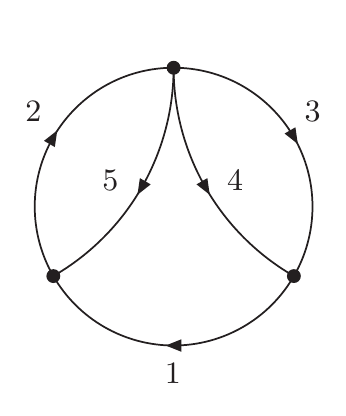} &
\includegraphics[scale=0.65]{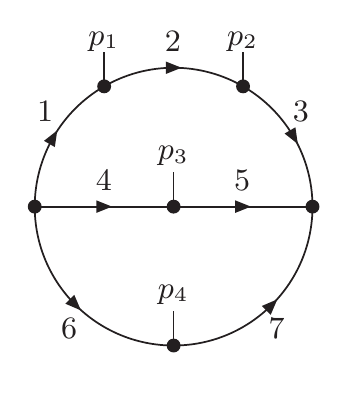}
\end{tabular}
\caption{Four graph examples to check loop-tree duality.}
\label{tab:graphs}
\end{table}
The integrals are defined as in Eq. (\ref{eq:intdef}) with $D=1$.
We note that the verification of our loop-tree duality approach in $D=1$ space-time dimensions
implies a verification in arbitrary space-time dimensions, hence it implies the equality of 
integrands of the spatial integrations.
The equality of the integrands means of course the equality of the integrals.
In order to avoid singularities (and the need of contour deformation) we chose the 
following masses for the propagators
\beq
\bsp
& m_1 = - 11 i \,,\quad m_2 = - 13 i \,,\quad m_3 = - 17 i \,,\quad m_4 = -23 i \,, \\
& m_5 = - 31 i \,,\quad m_6 = - 43 i \,,\quad m_7 = - 47 i \,.
\esp
\eeq
The external momenta for the non-planar double box was chosen in all outgoing kinematics as
\beq
p_1 = 1\,,\quad p_2 = 3 \,,\quad p_3 = 5 \,,\quad p_4 = - 9 \,.
\eeq
All four integrals were computed with the loop-tree duality method (LTD) and 
standard Monte Carlo integration (MC).
The results are presented in Table \ref{tab:compare} and the numbers show the agreement of 
the two methods.
For the sunrise and the non-planar double box we used the same combinatorial factors 
as their underlying chain graph is the same.
The results in Table \ref{tab:compare} confirm this statement as well.
\begin{table}[ht]
\centering
\renewcommand{\arraystretch}{1.4}
\begin{tabular}{|c|c|c|}
\hline
& Sunrise & Mercedes \\
\hline
$I_{LTD}$ & 
$\frac{1}{398684} \approx 2.50825 \cdot 10^{-6}$ & 
$\frac{3264791}{253676278437997615200} \approx 1.28699 \cdot 10^{-14}$ \\
\hline
$I_{MC}$ & 
$(2.5083 \pm 0.0001) \cdot 10^{-6}$ & 
$(1.28699 \pm 0.00008) \cdot 10^{-14}$\\
\hline
\hline
& Fly & N.p. double box\\
\hline
$I_{LTD}$ & 
$\frac{19}{653441364576} \approx 2.90768 \cdot 10^{-11}$ & 
$9.50190 \cdot 10^{-19}$ \\
\hline
$I_{MC}$ & 
$(2.9077 \pm 0.0002) \cdot 10^{-11}$ & 
$(9.504 \pm 0.005) \cdot 10^{-19}$\\
\hline
\end{tabular}
\caption{Graphs of Table \ref{tab:graphs} computed with loop-tree duality and direct Monte Carlo 
integration in $D=1$ space-time dimensions. The results agree at least up to 4 digits within uncertainty.}
\label{tab:compare}
\end{table}
%

%------------------------------------
%  Loop-tree duality for amplitudes 
%------------------------------------
%
\section{Loop-tree duality for amplitudes}
\label{sec:ltd-amps}
Now we move to the discussion of loop-tree duality for amplitudes.
For simplicity we use $\phi^3$ scalar theory to illustrate our approach and at the end of this section we 
explain the extension to gauge theories.

Our starting point is the $n$-leg $l$-loop renormalized scattering amplitude, which is the sum of the 
bare amplitude and the UV counterterms
\beq
A_{n,l} = A_{n,l}^{\text{bare}} + A_{n,l}^{\text{CT}} \,.
\eeq
We assume that UV counterterms have a local representation as integrals over the loop momenta.
This local representation matches the singularity structure of the bare amplitude 
and is a proper counterterm when integrated.
Thus formally we can write
\beq
\label{eq:amp}
A_{n,l} 
= \sum_{\Gamma \in U_{l,n}^{n.s.}} \frac{1}{S_{\Gamma}} 
\int \prod\limits_{i=1}^l \frac{\dd^D k_i}{(2 \pi)^D} f(\Gamma) \,,
\eeq
where $U_{l,n}^{n.s.}$ denotes the set of non-singular graphs 
with $n$ legs and $l$ loops and $S_{\Gamma}$ is the symmetry factor of the graph. 
In our notation here we do not differentiate between proper loop graphs and ones which 
correspond to local UV counterterms.
The interested reader can find the much more precise discussion in Ref. \cite{Runkel:2019zbm}.

Now we can apply the loop-tree duality theorem (\ref{eq:ltd-graphs}) for each graph in 
Eq. (\ref{eq:amp}) and obtain 
\beq
A_{n,l} =
(-i)^l \sum\limits_{\Gamma \in U_{l,n}^{n.s.}} \frac{1}{S_{\Gamma}}
\sum\limits_{\sigma \in C_{\Gamma}} \int\limits_{+/-} \prod\limits_{i=1}^l
\frac{\dd^{D-1} k_{\sigma_i}}{(2 \pi)^{D-1}}
S_{\sigma \alpha} (-1)^{n_{\sigma}^{(\alpha)}} \text{res} \Big(f,E_{\sigma}^{(\alpha)} \Big) \,.
\eeq
We relabel each loop momenta $k_{\sigma_i}$, average over all $l!$ possibilities and 
after relabeling we exchange the sum over cuts with the integration of loop momenta
\beq
\label{eq:ltd-amps1}
A_{n,l} =
\frac{(-i)^l}{l!} \int\limits_{+/-} \prod\limits_{i=1}^l
\frac{\dd^{D-1} k_i}{(2 \pi)^{D-1}}
\sum\limits_{\Gamma \in U_{l,n}^{n.s.}} 
\sum\limits_{\sigma \in C_{\Gamma}} 
\sum\limits_{S_l}
\frac{1}{S_{\Gamma}}
S_{\sigma \alpha} (-1)^{n_{\sigma}^{(\alpha)}} \text{res} \Big(f,E_{\sigma}^{(\alpha)} \Big) \,.
\eeq
As a result we obtain the renormalized loop amplitude as weighted sum of residues.
We would like to recognize this sum as an object, which can be computed more 
efficiently than just a sum of graphs.
To do so we have to address two issues: the presence of propagators with higher powers 
and symmetry factors.

We start with propagator with higher powers.
As we have seen in the previous section in Eq. (\ref{eq:residues}), residues 
can be recognized as tree graphs if all propagators occour to power one only.
However the presence of higher power propagators is unavoidable from two-loops and beyond.
The cuts of these propagators would introduce extra terms in the residues, which would destroy the 
tree structure and could introduce process dependence, which we would like to avoid.
In Ref. \cite{Baumeister:2019rmh} it was shown that at two-loops (and in principle at any loops) one can 
formulate local UV counterterms in such a way that they cancel the cuts of higher power propagators.
In this way the tree-like structure is kept, and the computation of higher power propagator cuts 
are reshuffled into process independent terms.

Now we discuss symmetry factors.
Symmetry factors are present for certain loop graphs, but are completely absent for any tree graph.
Since we have some combinatorial factors already one might argue that the symmetry factors could be 
absorbed into the combinatorial factors.
However symmetry factors depend on the graph itself, hence their absorption would introduce 
extra depencence for the combinatorial factors.
This would prevent the use recurrence relations in our method.
Hence we have to treat symmetry factors differently.
In Eq. (\ref{eq:ltd-amps1}) we have four different summations: we sum over all non-singular graphs 
with $n$ legs and $l$ loops, all possible $l$-cuts and all possible assignment of the $l$ loop 
momenta.
Finally there is an implicit summation over all energy configurations in the integration.
These four sums together are equivalent to a single sum over the set of non-singular graphs 
with $n$ legs and $l$ loops, where $l$ propagators are marked with an ordered $l$-tuple 
\beq
\int\limits_{+/-}
\sum\limits_{\Gamma \in U_{l,n}^{n.s.}} 
\sum\limits_{\sigma \in C_{\Gamma}} 
\sum\limits_{S_l} = \int \sum\limits_{U_{l,n}^{l-marked,n.s.}} \,.
\eeq
We name this set $U_{l,n}^{l-marked,n.s.}$. 
Eq. (76) of Ref. \cite{Runkel:2019zbm} allows us to replace the sum over the set $U_{l,n}^{l-marked,n.s.}$ 
with the set $U_{0,n+2l}^{l-sewed}$, the set of tree-graphs with $n+2l$ legs, where 
$l$ legs are sewed together.
The sum of the former includes symmetry factor, but the sum of the latter does not. 
The sewing operation of legs $k_i$ and $\bar{k}_i$ includes taking the limit 
$\bar{k}_i \rightarrow - k_i$. 

\begin{figure}
\centering
\includegraphics[scale=0.7]{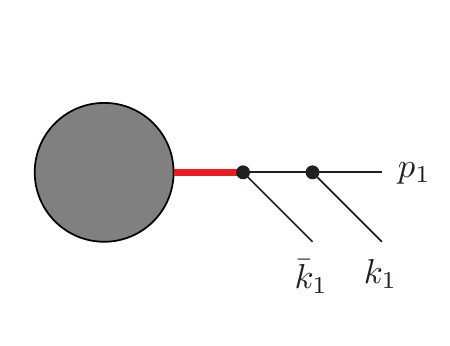}
\quad
\includegraphics[scale=0.7]{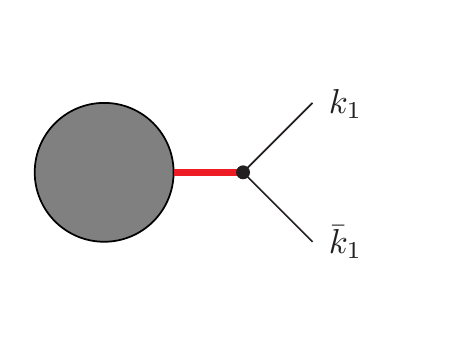}
\quad
\includegraphics[scale=0.7]{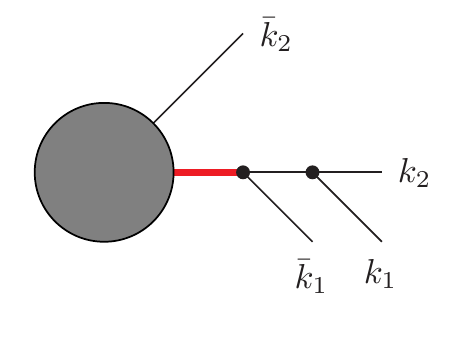}
\caption{Singular tree graphs in the forward limit. The vanishing inverse propagators are highlighted with red.}
\label{fig:singulargraphs}
\end{figure}

Using this relation we can change the summation of 
$l$-marked loop graphs into a summation of $l$-sewed tree graphs in Eq (\ref{eq:ltd-amps1}), obtaining 
\beq
\label{eq:ltd-amps2}
A_{n,l} =
\frac{(-i)^l}{l!} \int \prod\limits_{i=1}^l
\frac{\dd^{D-1} k_i}{(2 \pi)^{D-1} 2\sqrt{\vec{k}^2_i+m_i^2}}
\sum\limits_{\Gamma \in U_{0,n+2l}^{l-sewed,n.s.}} 
S_{\sigma \alpha} f(\Gamma) \,.
\eeq
On the right hand side we have a weighted sum of sewed tree graphs.
The question is if we can recognize this sum as an object, which can be computed 
efficiently.
As we mentioned the sewing operation includes taking the limit $\bar{k}_i \rightarrow - k_i$ 
for each sewed leg.
In case of tree amplitudes this would correspond to the $l$-fold forward limit.
In general the forward limit of tree amplitudes is singular due to the presence 
of vanishing inverse propagators.
These singular graphs are shown in Fig. \ref{fig:singulargraphs}.

\begin{figure}[ht]
\centering
\includegraphics[scale=0.6]{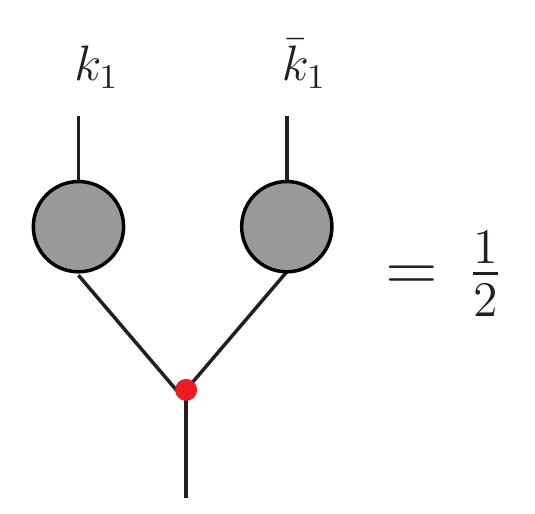}
\qquad \qquad
\includegraphics[scale=0.6]{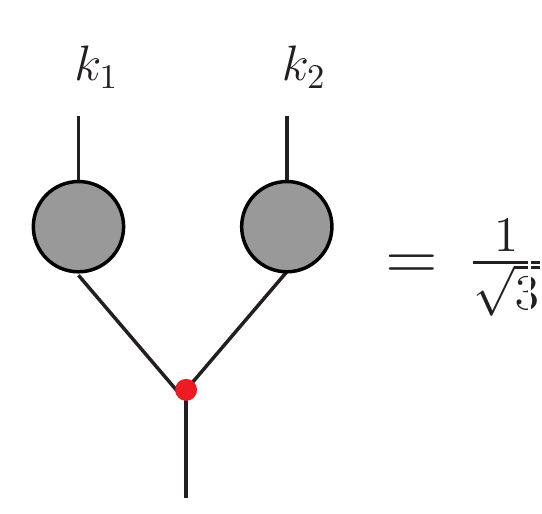}
\qquad \qquad
\includegraphics[scale=0.6]{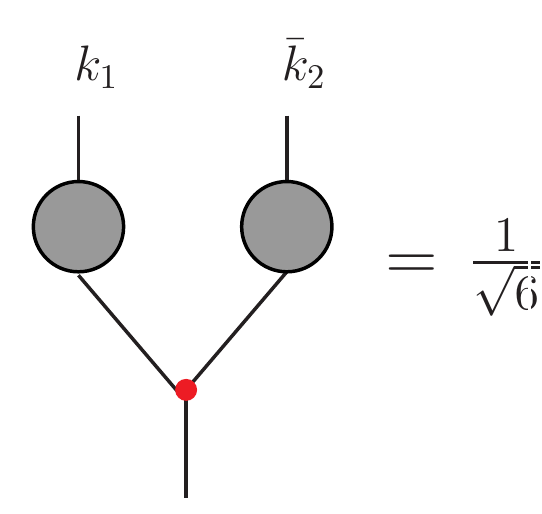}
\caption{Vertex rules for recurrence relations to incorporate combinatorial factors coming from cuts of  
two-loop graphs.}
\label{fig:combrules}
\end{figure}

The sewing of these type of tree-graphs correspond to self-energy insertion on external legs, 
tadpoles and self-energy insertions on internal lines resulting in higher power propagators.
These contributions are absent from the sum in Eq. (\ref{eq:ltd-amps2}) by construction.
Since we concern ourselves with scattering amplitudes, self-energy insertions on external legs 
are neglected from the beginning, since they would not contribute due to the LSZ reduction formula.
Tadpole diagrams can be ignored as well in theories with a vanishing vacuum expectation value.
Finally the third kind of diagrams, which originate from cuts of higher power propagators 
are cancelled by UV counterterms, hence they are not present as well.
Thus in Eq. (\ref{eq:ltd-amps2}) we have the regularized forward limit of the weighted sum of 
tree-graphs, what we name the tree-like object and is defined as
\beq
R_f \widetilde{A}_{0,n+2l}
= \lim_{\bar{k}_1 \rightarrow - k_1} \dots \lim_{\bar{k}_l \rightarrow - k_l}
\sum\limits_{\Gamma \in U_{0,n+2l}^{l-sewed,n.s.}}
S_{\sigma \alpha} f(\Gamma) \,,
\eeq
and then our loop-tree duality formula for renormalized loop amplitudes reads 
\beq
\label{eq:ltd-amps3}
A_{n,l} =
\frac{(-i)^l}{l!} \int \prod\limits_{i=1}^l
\frac{\dd^{D-1} k_i}{(2 \pi)^{D-1} 2\sqrt{\vec{k}^2_i+m_i^2}} 
R_f \widetilde{A}_{0,n+2l} \,.
\eeq
It is important to note that here we do not refer to individual Feynman-diagrams anymore.
$R_f \widetilde{A}_{0,n+2l}$ has a tree structure identical to the 
regularized forward limit of a proper tree amplitude with the same number of external legs.
Tree amplitudes can be computed efficiently via off-shell recurrence relations 
\cite{Berends:1987me,Kosower:1989xy,Draggiotis:2002hm,Weinzierl:2005dd,Dinsdale:2006sq,Duhr:2006iq}.
For the tree-like object we can modify the rules of recurrence relations 
to incorporate the presence of combinatorial factors, which was presented up to three-loops 
in Ref. \cite{Runkel:2019zbm}.
For example at two-loops we can assing factors to vertices, whose value depends on the 
sewed loop momenta entering the vertex, as illustrated on Fig. \ref{fig:combrules}.

In order to check that the set of graphs matches on both sides of Eq. (\ref{eq:ltd-amps3}), 
we performed the following calculation: on one hand we generated every loop graph (including their symmetry factor) 
of the $n$-leg $l$-loop bare amplitude using \texttt{QGRAF} \cite{Nogueira:1991ex}.
We performed all possible cuts to obtain the set of tree graphs and removed those, which correspond 
to cuts of propagators with higher powers.
We searched for identical graphs, and added them by combining their symmetry factor. 
We verified that all symmetry factors cancel.
On the other hand we generated all corresponding tree graphs using recurrence relations. 
Here we filtered out singular graphs, which correspond to configurations shown in 
Fig. \ref{fig:singulargraphs}.
At last we compared the lists of tree graphs obtained in these two different ways, and verified 
that the two lists are in complete agreement.
We performed this check for every combination of $n$ and $l$ such that $n+2l=9$, and also $l \le 3$, 
the three-loop case containing $\ordo{10^5}$ graphs.

Finally we discuss the extension of our method from scalar theory to gauge theories.
In theories with different particle flavors we have to sum over the flavor of the 
sewed legs.
In Feynman-gauge this includes ghost particles as well.
Sewing two external legs must include summation over both physical and unphysical spin states and 
further degrees of freedom as well, for example color.
Additionaly, the sewing of fermion or ghost legs must include a factor of $(-1)$ factor in order 
to reproduce the correct loop factor.

%------------------------------------
%             Summary 
%------------------------------------
%
\section{Summary}
\label{sec:summary}
In this talk we presented a new formulation of the loop-tree duality method.
We showed how single graphs with an arbitrary 
number of loops can be related to a weighted sum of tree graphs.
Using this relation we showed that renormalized scattering amplitudes 
with $l$ loops can be expressed in terms of tree-like objects.
We illustrated that these tree-like objects can be computed with the 
help of off-shell recurrence relations with a modified set of rules, which 
incorporate the presence of combinatorial factors.
Our method is general and does not rely on any information of the process in question, 
hence it can serve as a basis of computer program to calculate higher loop amplitudes 
numerically with an arbitrary number of legs.
This and the possibility of local cancellation of infrared singularities will be explored 
in future publications.

%------------------------------------
%          Bibliography 
%------------------------------------
%
\providecommand{\href}[2]{#2}\begingroup\raggedright\endgroup
\end{document}